\documentclass[prb,twocolumn]{revtex4-1} 

\usepackage{amsmath}  
\usepackage{amsfonts} 
\usepackage{graphicx} 

\begin{document}


\title{Dynamics of Trees of Fragmenting Granules in the Quiet Sun: \textit{Hinode}/SOT Observations Compared to Numerical Simulation}

\author{J.M. Malherbe} \email{Jean-Marie.Malherbe@obspm.fr}
\affiliation{LESIA, Observatoire de Paris, 92195 Meudon,France}

\author{T. Roudier} \email{Thierry.Roudier@irap.omp.eu}
\affiliation{IRAP, Universit\'{e} de Toulouse, CNRS, CNES, 14 Avenue Edouard Belin, 31400 Toulouse,
              France}
\author{R. Stein} \email{Stein@pa.msu.edu}
\affiliation{Physics and Astronomy Department, Michigan State University, East Lansing, MI 48824, USA}

\author{Z. Frank} \email{Zoe@lmsal.com}
\affiliation{Lockheed Martin Solar and Astrophysics Laboratory, Palo Alto, 3251 Hanover Street, CA 94303,
              USA}


\date{\today}

\begin{abstract}

We compare horizontal velocities, vertical magnetic fields and
evolution of trees of fragmenting granules (TFG, also named families
of granules) derived in the quiet Sun at disk center from
observations at solar minimum and maximum of the \emph{Solar Optical
Telescope} (SOT on board \textit{Hinode}) and results of a recent 3D
numerical simulation of the magneto-convection. We used 24-hour
sequences of a 2D field of view (FOV) with high spatial and temporal
resolution recorded by the SOT \emph{Broad band Filter Imager} (BFI)
and \emph{Narrow band Filter Imager} (NFI). TFG were evidenced by
segmentation and labeling of continuum intensities. Horizontal
velocities were obtained from local correlation tracking (LCT) of
proper motions of granules. Stokes \emph{V} provided a proxy of the
qline of sight magnetic field (BLOS). The MHD simulation (performed
independently) produced granulation intensities, velocity and
magnetic field vectors. We discovered that TFG also form in the
simulation and show that it is able to reproduce the main properties
of solar TFG: lifetime and size, associated horizontal motions,
corks and diffusive index are close to observations. Largest (but
not numerous) families are related in both cases to strongest flows
and could play a major role in supergranule and magnetic network
formation. We found that observations do not reveal any significant
variation of TFG between solar minimum and maximum.

  \end{abstract}

\maketitle 

\section{Introduction}
Roudier \emph{et al.} (2016), referenced below as paper I, found
from \textit{Hinode} observations a relationship between the
evolution of trees of fragmenting granules (TFG) and horizontal
motions of magnetic features in the quiet Sun. They suggested that
TFG could contribute to the formation of the magnetic network.

Reviews by Sheeley (2005) and Ossendrijver (2003) suggest that the
quiet Sun contributes significantly to the solar magnetism.
Understanding how magnetic fields form in the convective zone and
are advected and diffused to the surface is a challenge which
requires to investigate flows at various length scales (granulation,
meso- and super-granulation), both from observations and MHD
simulation (see review by Stein, 2012).

The solar granulation is organized in families of granules (TFG),
which can live several hours. They are formed by successively
exploding granules originating from a single parent (for the history
of the discovery, please refer to paper I and Roudier \emph{et al.},
2003). We suspect TFG to be implied in network formation through
mesoscale surface flows. Observations and simulations at similar
space and time resolutions are necessary to characterize and
understand their impact on magnetic elements.

In the present paper, the goal is to compare several \textit{Hinode}
observations at different dates along the cycle with an independent
numerical simulation to examine how realistic it is in terms of TFG
formation, evolution and interaction with magnetic elements. For
this purpose, we analyze results coming out from several 24-hour
sequences observed with \textit{Hinode} (near solar minimum in 2007
and maximum in 2013, 2014) and the recent (2014) numerical
simulation of magneto-convection based on Stein and Nordlund (1998).
We study the dynamics of the solar surface at the mesoscale,
relationships between TFG and horizontal motions, spatial power
spectra, TFG size and lifetime, corks and diffusive index. We also
investigate the action of TFG in the formation of the magnetic
network.

\section{Description of Observations and Processing Methods}

We used multi-wavelength data sets of the Solar Optical Telescope
(SOT) on board \textit{Hinode} (for a description of the SOT, see
Ichimoto \textit{et al.}, 2005; Suematsu \textit{et al.}, 2008). The
0.50 m aperture of the SOT provides a spatial resolution of about
0.25$''$ (180 km) at 450 nm. Observations were done in the frame of
HOP217 (determination of the properties of families of granules and
formation of the photospheric network) and HOP295 (connection of
families of granules to the formation of the magnetic network). We
observed at disk center, so that horizontal velocities can be
derived from the proper motions of granules. Families can be
detected from time evolution of granules.

For horizontal velocity measurements and TFG analysis, the BFI
provided 24-hour sequences of the blue continuum at 450.4 nm.
Observations were recorded continuously on 29-30 August 2007, from
10:48 to 10:40 UT and 30-31 August 2007 from 10:50 to 10:18 UT (two
consecutive sequences, 2007a and 2007b, which can be combined to
form a long 48 hours sequence). It must be noticed that the 2007
sequence is not far from a small remnant activity located south
east. We also selected the 24-hour sequences of 28-29 April 2013,
from 06:13 to 06:13 UT and 28-29 December 2014, from 22:14 to 22:56
UT (table).

After alignment of data (which reduced slightly the initial FOV), we
applied a subsonic Fourier filter in the $k-\omega$ space, where $k$
and $\omega$ are respectively the horizontal wave number and
pulsation, in order to remove oscillations of continuum intensities.
All Fourier components such that $ \omega \geq C_{\textrm{s}} k$
(where $C_{\textrm{s}}$ = 6 km s$^{-1}$ is the sound speed and $ k =
\sqrt{k_{x}^{2}+k_{y}^{2}}$ is the horizontal wave number) were
retained to keep only convective motions (Title \emph{et al.},
1989). In order to analyse TFG, we detected exploding granules and
labeled them in time after segmentation as described by Roudier
\emph{et al.} (2003). Families or their different branches
correspond to the mesogranular scale.

Horizontal velocities ($ v_{x} $ and $ v_{y} $) were also derived
from aligned and filtered data of the blue continuum using the LCT
technique (November and Simon, 1988, Roudier \emph{et al.}, 1999)
with a temporal window of 30 minutes and a spatial window of
$3.5''$, providing mean motions at the mesoscale or larger. As such
flows evolve slowly and families need at least 6 hours to develop,
this explains why we worked with 24-hour sequences. The LCT is based
on the detection of spatial and time-averaged granular proper
motions.For vertical magnetic fields (BLOS), we used the NFI in shuttered
Stokes $I$ and $V$ mode to record the $V$ signal (a proxy of the
magnetic field) in the blue wing of Fe \textsc{i} 630.2 nm (exposure
time 0.12 seconds). The observations were recorded together with the
BFI, at the same time and cadence. The Lyot filter was tuned to a
single wavelength 12 pm apart from line core. $I+V$ and $I-V$ images
were taken, then added or subtracted (providing respectively Stokes
$I$ and $V$) on board the spacecraft. We also selected the long
sequence of very quiet Sun in Na \textsc{i} D1 589.6 nm (shuttered
$V/I$ mode, exposure time 0.20 seconds) from 30 December 2008 (10:30
UT) to 5 January 2009 (05:37 UT) with 5 minutes time step which has
a better sensitivity and a much lower JPEG type compression noise
(table).

\begin{table}
\begin{tabular}{|c|c|c|c|c|c|c|}
  \hline
  Instrument & $\lambda$ & Time & Pixel & Duration & FOV &
  Usage \\
   &(nm) & step (s) & ($''$) &(hours) & ($''$) & \\
  \hline
  BFI 2007 & 450.4 & 50.2 & 0.11 & 24 & 112 $\times$ 112 & LCT+TFG \\
  BFI 2013 & 450.4 & 40.0 & 0.11 & 24 & 84 $\times$ 89 & LCT+TFG \\
  BFI 2014 & 450.4 & 60.0 & 0.11 & 24 & 77 $\times$ 77 & LCT+TFG \\
  IRIS 2015 & 283.2 & 60.0 & 0.17 & 6 & 60 $\times$ 60 & LCT \\
  NFI 2007 & 630.2 & 50.2 & 0.16 & 24 & 112 $\times$ 112 & BLOS \\
  NFI 2008 & 589.6 & 300 & 0.16 & 120 & 112 $\times$ 112 & BLOS \\
  \hline
\end{tabular}
\caption{List of observations}
\end{table}

\section{Description of the MHD Simulation}

We used the results of the 3D magneto-convection stagger code which
was run independently and not designed to model solar TFG. It solves
the equations of mass, momentum and internal energy in conservative
form plus the induction equation of the magnetic field, for
compressible flow on a staggered mesh (Stein and Nordlund, 1998;
Stein \emph{et al.}, 2009; review by Stein, 2012). Solar rotation is
included. Scalar variables (density, internal energy and
temperature) are volume centered, momenta and magnetic field are
face-centered, while currents and electric field are edge-centered.
Time integration is performed by a 3rd order low memory Runge-Kutta
scheme. Parallelization is achieved with MPI, communicating the
three overlap zones that are needed in the 6th and 5th order
derivative and interpolation stencils.

Boundaries are periodic horizontally and open at the top and bottom.
The code uses a tabular equation of state that includes local
thermodynamic equilibrium (LTE) ionization of the abundant elements
as well as hydrogen molecule formation, to obtain the pressure and
temperature as a function of log density and internal energy per
unit mass. Radiative heating/cooling is calculated by explicitly
solving the radiation transfer equation in both continua and lines
assuming LTE. The number of wavelengths for which the transfer
equation is solved is drastically reduced by using a multi-group
method which accurately models the photospheric structure as
revealed in line profiles and limb darkening.

Because the stagger code includes all the significant physical
processes occurring near the solar surface and resolves the thin
thermal boundary layer at the top of the convection zone, its
results can be used to make predictions and be compared
quantitatively with various solar observations, as oscillations
(Rosenthal \emph{et al.}, 1999; Stein and Nordlund, 2001), Fe
\textsc{i} line formation (Asplund \emph{et al.}, 2000), G band
magneto-convection (Carlsson \emph{et al.}, 2004),  facular
variability (De Pontieu \emph{et al.}, 2006), velocities of magnetic
elements (Langangen \emph{et al.}, 2007), heliosismology (Zhao
\emph{et al.}, 2007) or correlation tracking (Georgobiani \emph{et
al.}, 2007).
The initial state was a snapshot from a non-magnetic convective
simulation.  This in turn was the result of a series of hydrodynamic
convection runs.  All had the same 20 Mm depth.  A 12 Mm wide
simulation was run for two turnover times; then it was doubled three
times to 96 Mm wide before the magnetic field was introduced (after
37.65 hours of hydrodynamic convection) as a 100 G uniform
horizontal field, advected by inflows into the computational domain
at the bottom.  It took about 2 days (one turnover time) for
significant magnetic flux to reach the surface.

Runs have dimensions 2016 $\times$ 2016 $\times$ 500 with resolution
48 km horizontally and 12-80 km vertically. Quantities at the
visible surface were binned 2 $\times$ 2 from 48 to 96 km (0.13$''$
pixel size) to be comparable to observations (0.11$''$). The
physical parameters issued from the simulation are the emergent
intensity, velocity field and magnetic field vectors (we used only
horizontal velocities and the BLOS component). The FOV was 131$''$
$\times$ 131$''$ and time step 60 seconds. The run duration covered
a long time (100 hours), but we extracted a 26 hours sequence (start
time: 59.5 hours; end time: 85.5 hours) for comparison to
observations.

Several MP4 movies are joined as Electronic Supplemental Material
(see the Appendix for details).

\section{Horizontal Velocities of Simulation: Comparison Between Plasma, LCT, FLCT and CST}
\begin{figure}
\centering
\includegraphics[width=8 cm]{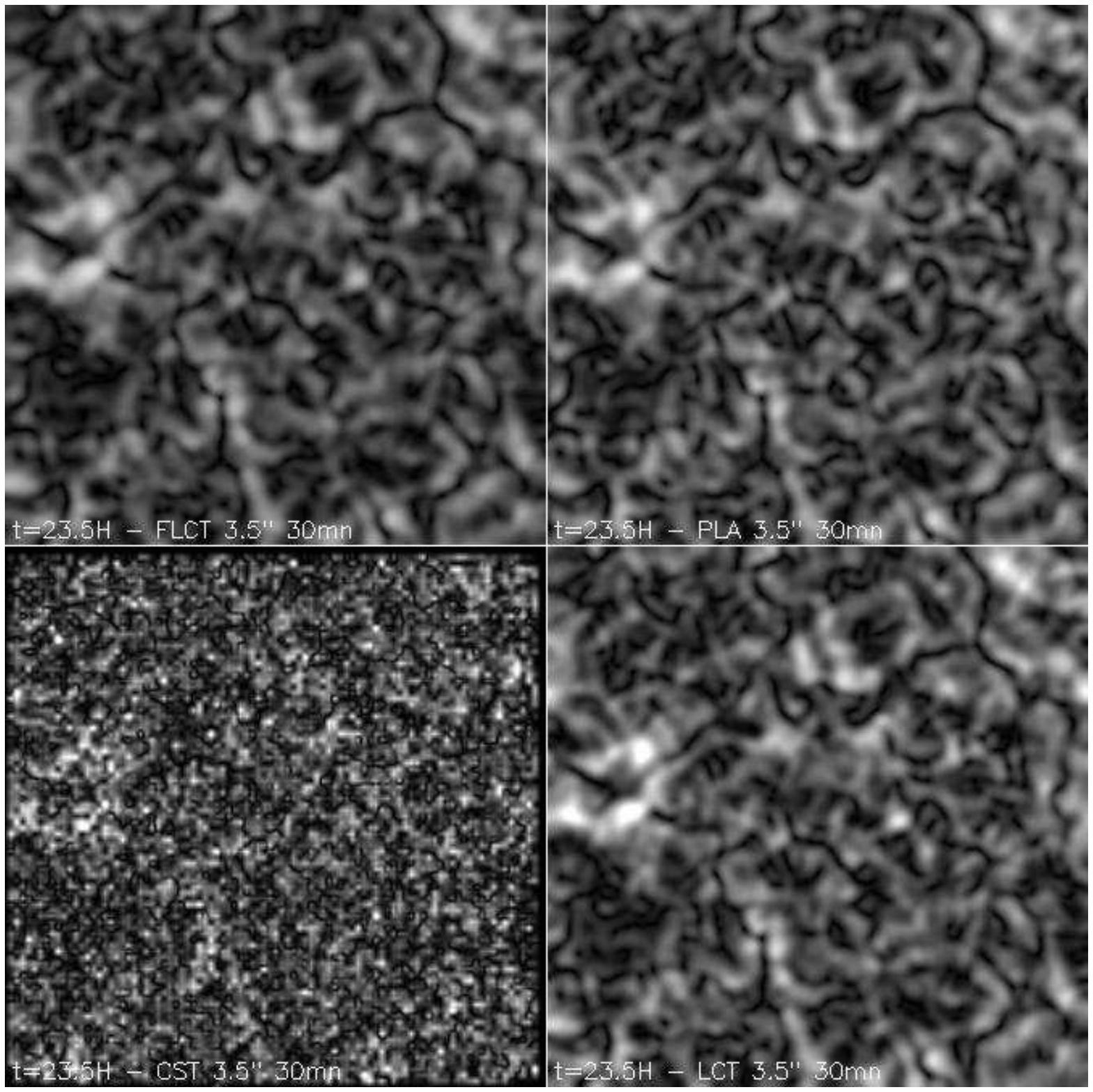}
 \caption{Comparison between plasma, LCT, FLCT and CST horizontal
 velocity module using an averaging 30 minutes temporal window and
 3.5$''$ spatial window.
 FOV = $131'' \times 131''$. Top left: FLCT velocity; top right: plasma velocity;
 bottom left: CST velocity; bottom right: LCT velocity} \label{compar}
\end{figure}

The numerical simulation provides the horizontal velocity vector
over the 2D FOV during 26 hours with 60 seconds time step. It is a
good opportunity to compare the plasma velocity to the one issued
from indirect methods, such as LCT, FLCT (Fourier LCT, Fisher and
Welsch, 2008) or CST (Coherent Structure Tracking, Rieutord \emph{et
al.}, 2007), applied to intensity structures as granules. FLCT
compares two consecutive images while LCT and CST are time averaged.
We used 30 minutes and $3.5''$ averaging windows. Indeed Rieutord
\emph{et al.} (2001) showed that granules are good tracers of solar
surface velocities for space and time scales not shorter than these
typical values. In order to compare the quality of structure
tracking to plasma motion, we applied the same windows to horizontal
flows. Figure~\ref{compar} shows that the correlation between plasma
and LCT or FLCT velocities is morphologically good; this is not the
case of the CST (which gives, on the contrary, much better results
at $1''$ resolution, see Roudier \emph{et al.}, 2012). The next two
figures provide quantitative results.

\begin{figure}
\centering
\includegraphics[width=8 cm ]{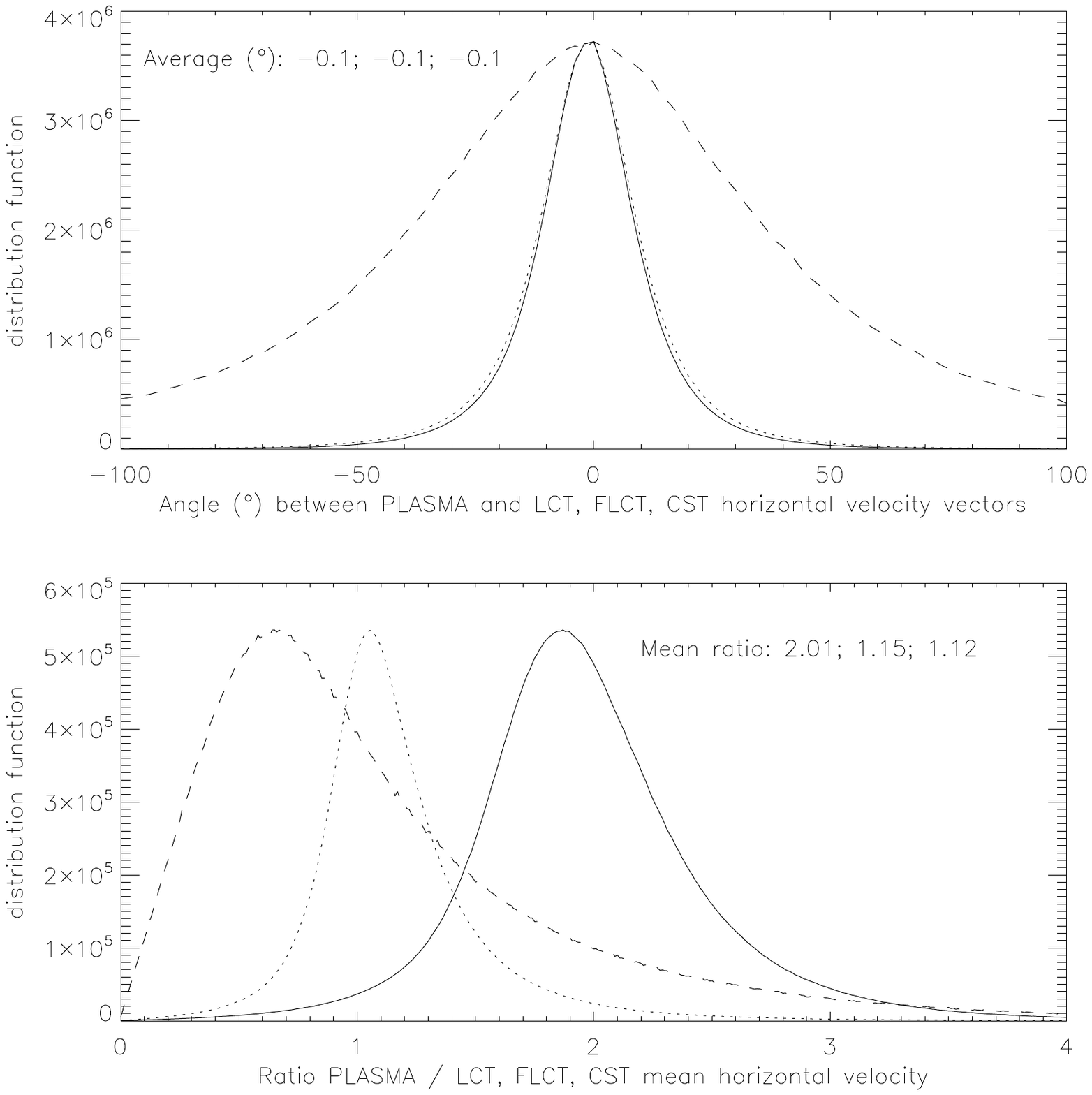}
 \caption{Comparison of PLA, LCT, FLCT and CST horizontal velocities of simulation.
 Top: mean angle between velocity vectors (solid: PLA, LCT; dotted: PLA, FLCT; dashed: PLA, CST).
 Bottom: mean ratio between velocity modules (solid: PLA/LCT; dotted: PLA/FLCT; dashed: PLA/CST). Average
 values are indicated for direction and ratio.} \label{histo}
\end{figure}

Figure~\ref{histo} allows to compare performances of LCT, FLCT and
CST for windows of 30 minutes/3.5$''$. We analysed the departures
between the direction of horizontal velocities given by these
methods and the plasma velocity vector (PLA), filtered by the same
windows. The mean angle between LCT, FLCT, CST and PLA is almost
null, but the dispersion is 3.75 times higher with the CST (standard
deviation 34 degrees instead of 9 degrees for LCT or FLCT); the mean
ratio between PLA and LCT velocity module is 2.0; while between PLA
and CST, or PLA and FLCT, we got respectively 1.15 and 1.12. The
distribution function is definitely asymmetric for CST with an
important tail. The best method is FLCT, but as it is slow for long
sequences and large images, so we used the LCT instead to determine
mesoscale horizontal flows, even if it underestimates the velocity
module.

\section{Horizontal Velocities from LCT: Observations and Simulation}

We now compare horizontal velocities provided by the LCT applied to
\textit{Hinode} observations and simulation. We also add a new
Interface Region Imaging Spectrograph observation (IRIS, HOP312),
performed on 10-11 October 2015 (23:34 to 05:32 UT) at disk center
with the slit jaw (ultra violet continuum at 283.2 nm) with 13.8
seconds time step, 0.166$''$ pixel size and 6 hours duration
sequence; the IRIS, \textit{Hinode} and simulation data sets do not
differ much in pixel size. We used the same temporal and spatial
windows for all (30 minutes, 3.5$''$).

\begin{figure}
\centering
\includegraphics[width=8 cm]{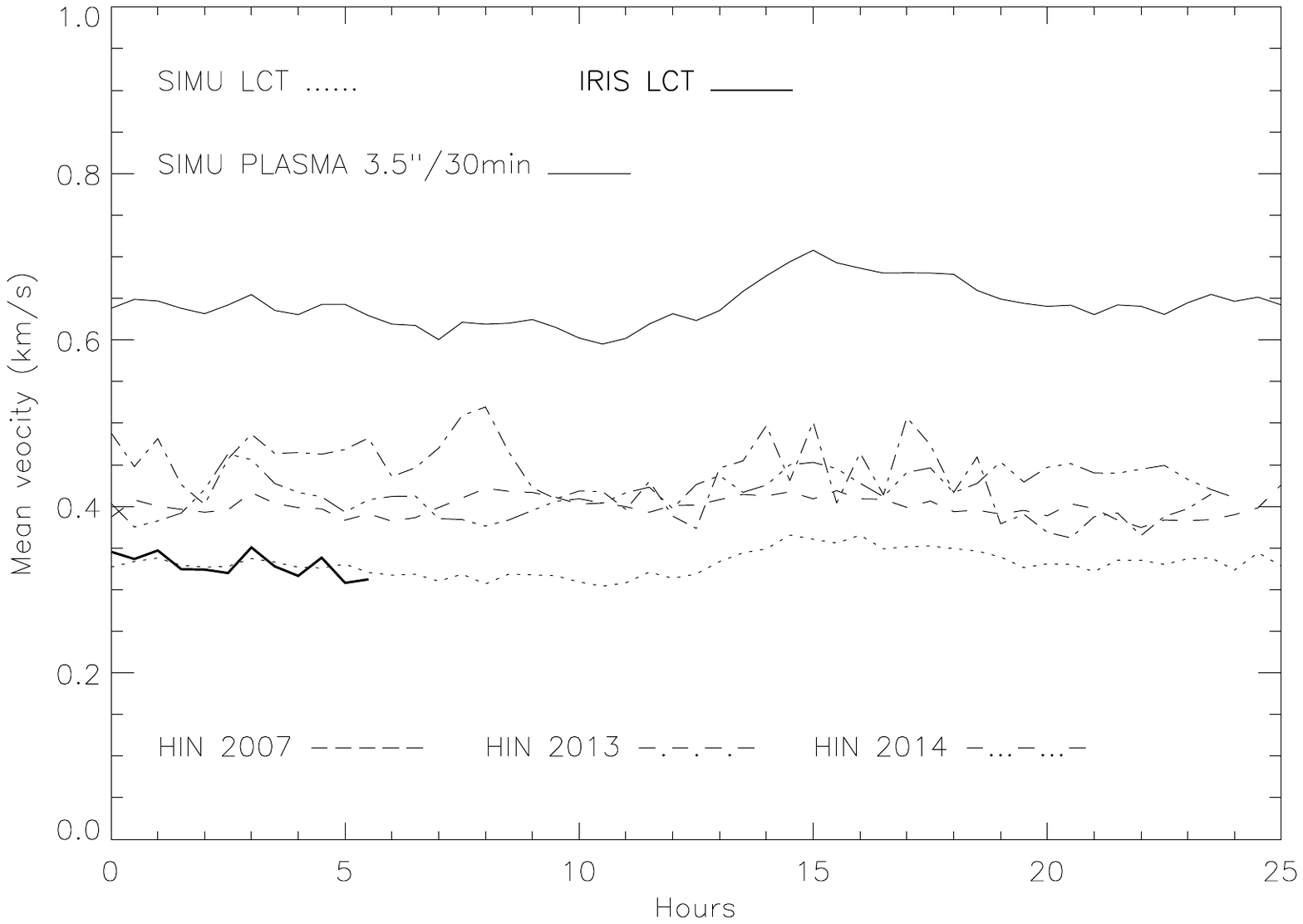}
 \caption[]{Mean horizontal velocity over the FOV as a function of time.
 Solid line: plasma velocity (simulation);
dotted line: LCT (simulation); dashed line: LCT \textit{Hinode}
2007; dash-dot: LCT \textit{Hinode} 2013; dash-dot-dot: LCT
\textit{Hinode} 2014; thick solid line: LCT IRIS 2015.
 } \label{vstat}
\end{figure}

Figure~\ref{vstat} shows, as a function of time, the behavior of the
plasma velocity module and LCT proxy of the simulation, together
with LCT velocity modules of \textit{Hinode} (2007, 2013, 2014) and
IRIS (2015). Velocities are averaged on the FOV and remain almost
constant in time. If we apply a factor 2.0 to LCT observations, we
find a mean horizontal velocity of 0.8 km/s, close to the 0.65-0.70
km/s of the simulation.

\section{Magnetic Fields of the Quiet Sun}

\begin{figure}
\centering
\includegraphics[width=8 cm]{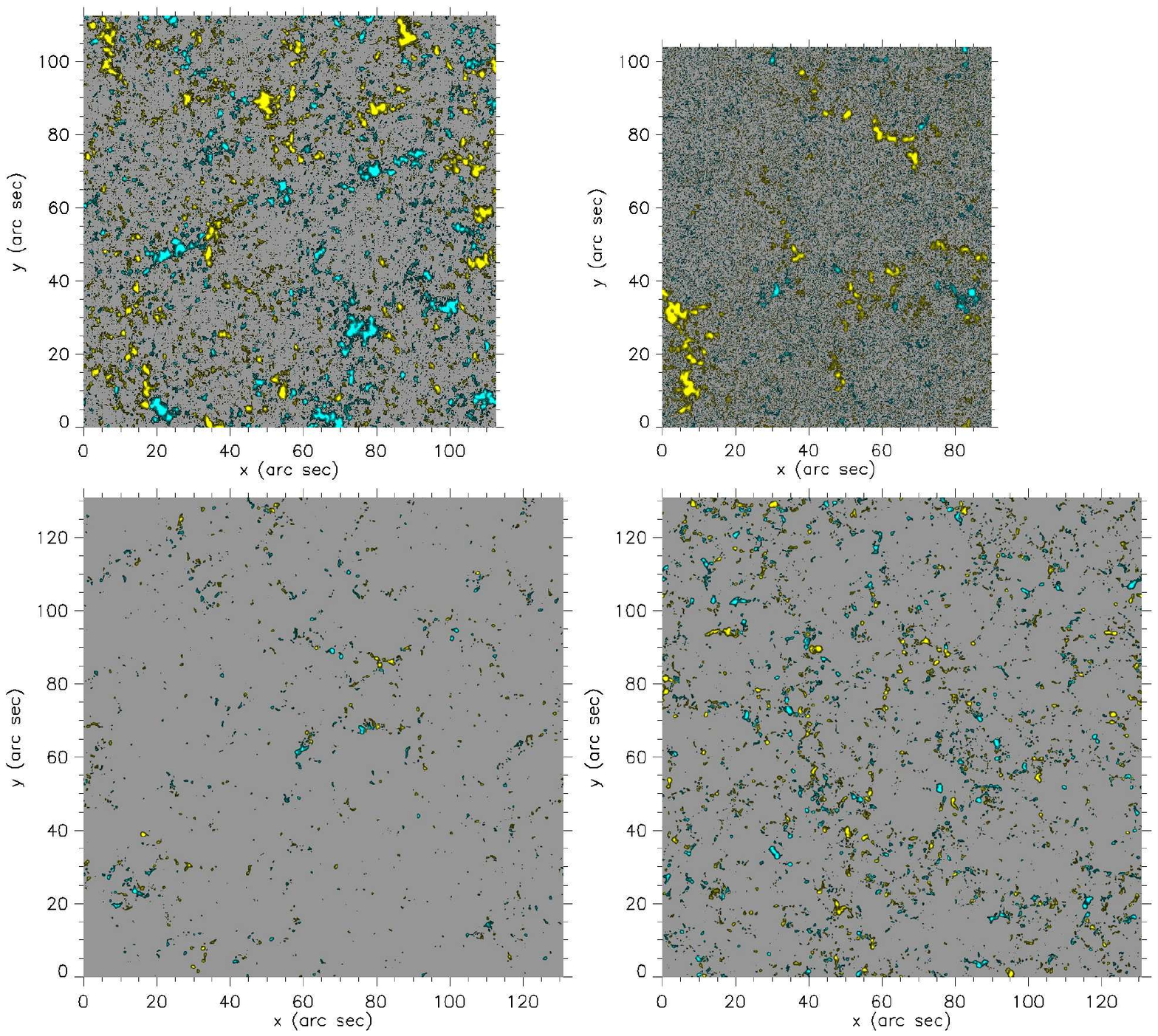}
 \caption{Vertical magnetic field (blue/yellow for north/south polarities);
top left: \textit{Hinode} 2008; top right: \textit{Hinode} 2007;
bottom left: simulation start at 59.5 hours; bottom right:
simulation end at 85.5 hours, filtered by the SOT PSF.
 } \label{bmag}
\end{figure}

Figure~\ref{bmag} shows typical magnetic fields (BLOS) of
\textit{Hinode} 2007 and 2008 at disk center (0.16$''$ pixel size),
together with the beginning and the end of the 26 hours numerical
sequence (0.13$''$ pixel) extracted from the full run. The 2008 FOV
shows several supergranular cells, but also that there exists a lot
of spatially concentrated and mixed polarities inside cells, as
noticed by Dominguez Cerde\~{n}a \emph{et al.} (2006). The 2007 FOV
is more noisy and contains a well structured supergranule with BLOS
mainly South. The numerical simulation is not magnetically steady
state (there is a constant input of horizontal field at the bottom
of the box); at the surface, there is more field at the end of the
sequence than at the beginning. Polarities are much more mixed than
in observations, with narrow concentrated flux tubes, so that it is
not so easy to distinguish the network. As the spatial resolution of
BLOS is much better in the simulation, we filtered it by the point
spread function (PSF) of the SOT (including the 50 cm aperture,
central occultation, spider and CCD, assuming no defocus). Orozco
Su\'{a}rez \emph{et al.} (2007) pointed out the role of instrumental
degradation for imaging magnetic fields with \emph{Hinode}. The 2008
sequence (with the best magnetic sensitivity) exhibits a spatial
distribution fairly close to the simulation, with well formed
network and mixed polarities inside.

\begin{figure}
\centering
\includegraphics[width=8 cm]{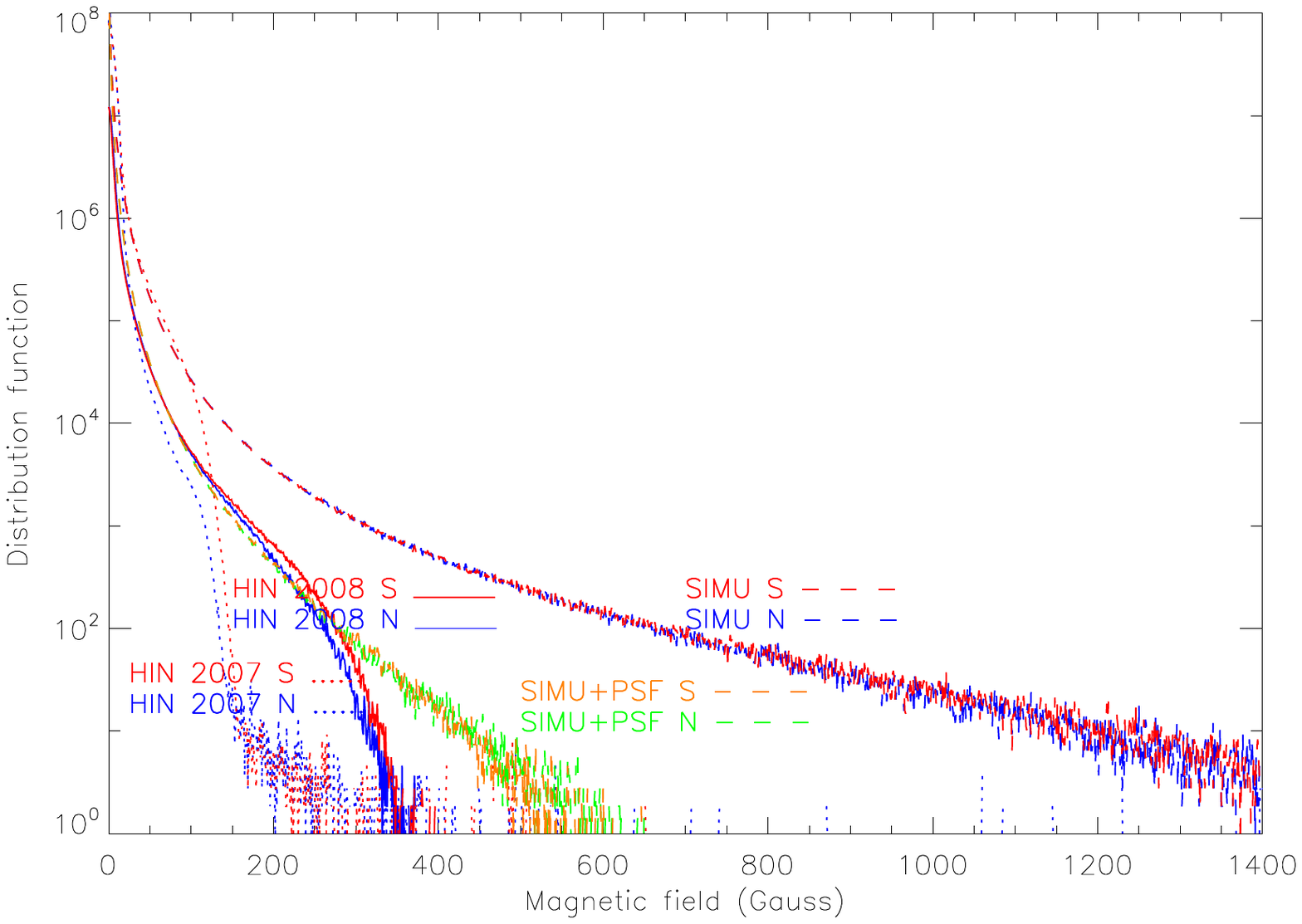}
 \caption{Distribution functions of magnetic polarities (blue/red for north/south polarities);
 solid line: \textit{Hinode} 2008, Na \textsc{i} D1; dotted line: \textit{Hinode} 2007, Fe \textsc{i}; dashed line: simulation.
 The green/yellow curves (North/South) are for the
 simulation filtered by the SOT PSF.
 } \label{bmagdis}
\end{figure}

The distribution functions of the vertical magnetic field
(Figure~\ref{bmagdis}) exhibit an asymmetry between north and south
polarities for \textit{Hinode} 2007 (probably due to the vicinity of
a weak active region). On the contrary, there is almost no asymmetry
in 2008 (very quiet Sun region); polarities of the numerical
simulation are well balanced but produce concentrated fields much
more intense. In particular, BLOS is smaller than 200 Gauss for
\textit{Hinode} 2007 (350 Gauss for 2008), while kilo Gauss (kG)
fields are produced by the MHD code. However, Figure~\ref{bmagdis}
shows that kG fields vanish when the SOT PSF is applied to the
simulation. BLOS is likely underestimated by the Stokes sensitivity
and the spatial resolution of the SOT and NFI filter.

In particular, the simulation reveals the presence of mainly
vertical kG flux tubes only in intergranular lanes (median angle
with the vertical 9 degrees, 90\% of angles smaller than 17
degrees). The vertical velocity in flux tubes is downward ($-2.23$
km/s in average). The downflow increases with magnetic field
strength ($-1.92$, $-2.67$ and $-3.43$ km/s respectively for fields
above $1.0$, $1.5$ and $2.0$ kG). 97\% of flux ropes produce
downdrafts. Unfortunately, the spatial resolution of \textit{Hinode}
does not allow to check these predictions.

\section{Spatial Power Spectra of Horizontal Velocities}

\begin{figure}
\centering
\includegraphics[width=8 cm]{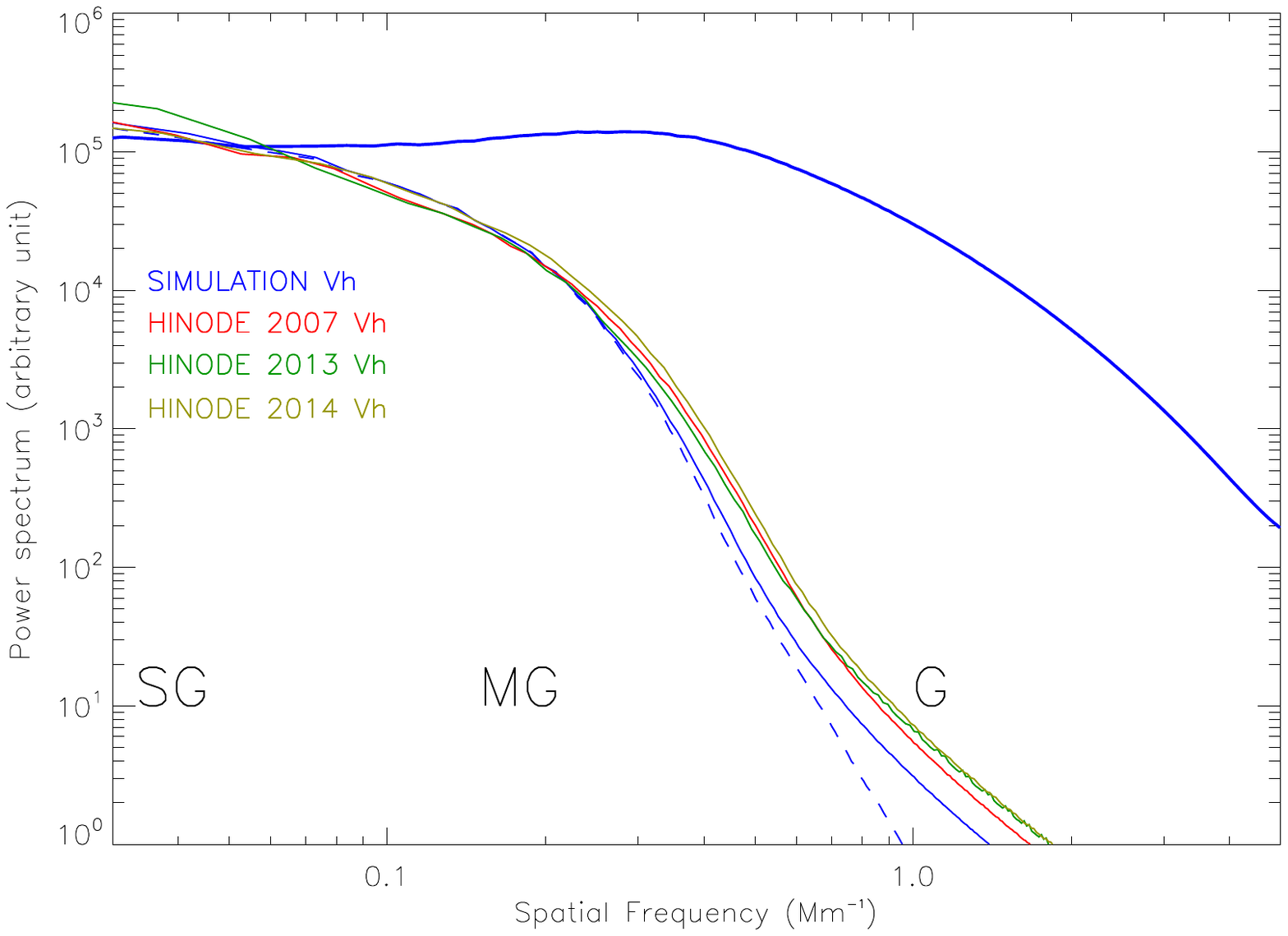}
 \caption{Power spectra of horizontal velocities of observations
 and simulation. LCT velocities of observations: 2007 (red), 2013 (green), 2014 (dark green).
 Horizontal velocities of the simulation (blue): plasma velocity (thick), LCT velocity (thin)
 and plasma velocity after convolution by LCT windows (dashed).
  SG, MG and G designate typical supergranular, mesogranular
  and granular length scales.} \label{spec}
\end{figure}

Figure~\ref{spec} provides the power spectra of horizontal
velocities of \textit{Hinode}/BFI observations as well as the
simulation. The power spectrum is defined as the square module of
the 2D spatial Fourier transform averaged over the sequence duration
and integrated over circular coronas. Power spectra of horizontal
velocities computed from the LCT are in full agreement at the
supergranular (30 Mm) and mesogranular (5 Mm) scales for
observations and simulation. The power spectrum of the plasma
velocity of the simulation (convolved by the temporal and spatial
LCT windows) also agrees with LCT results. The filtering windows
make the power spectra of horizontal velocities vanish at the
granular scale (1 Mm); of course, this is not the case for plasma
velocities of the simulation.

\section{Granule Merging and Splitting Rates}

Splitting or exploding granules produce mesoscale horizontal flows
and form families. We determined the splitting rate (fraction of
exploding granules) and merging rate (fraction of collapsing
granules) in BFI observations and simulation as a function of time.
We did not find any significant time variation during the 24-hour
sequences: the merging rate is 0.065 for the simulation (0.11 for
\textit{Hinode} 2007). The splitting rates are always higher: 0.115
for the simulation (0.14 for \textit{Hinode} 2007). Splitting rates
are always higher than merging rates; this explain the formation of
TFG which appear both in observation and simulation and will be
analysed in the next section.

\section{Dynamics of Trees of Fragmenting Granules (TFG)}

In paper I, we described the main properties of TFG. Their areas
grow by a succession of granule explosions (some can occur
simultaneously). TFG are able to push magnetic elements of the
internetwork (IN) towards the frontiers of supergranules to form the
network (NE). Explosions of granules at the TFG birth generate flows
at the mesogranular scale in the range 0.5 to 0.8 km/s, which
propagate outwards when there is no counterpart flow of adjacent
families. We also evidenced the existence of large velocity fronts
affecting the NE shape. Magnetic features are often located at the
edge of such fronts.

\begin{figure}
\centering
\includegraphics[width=8 cm]{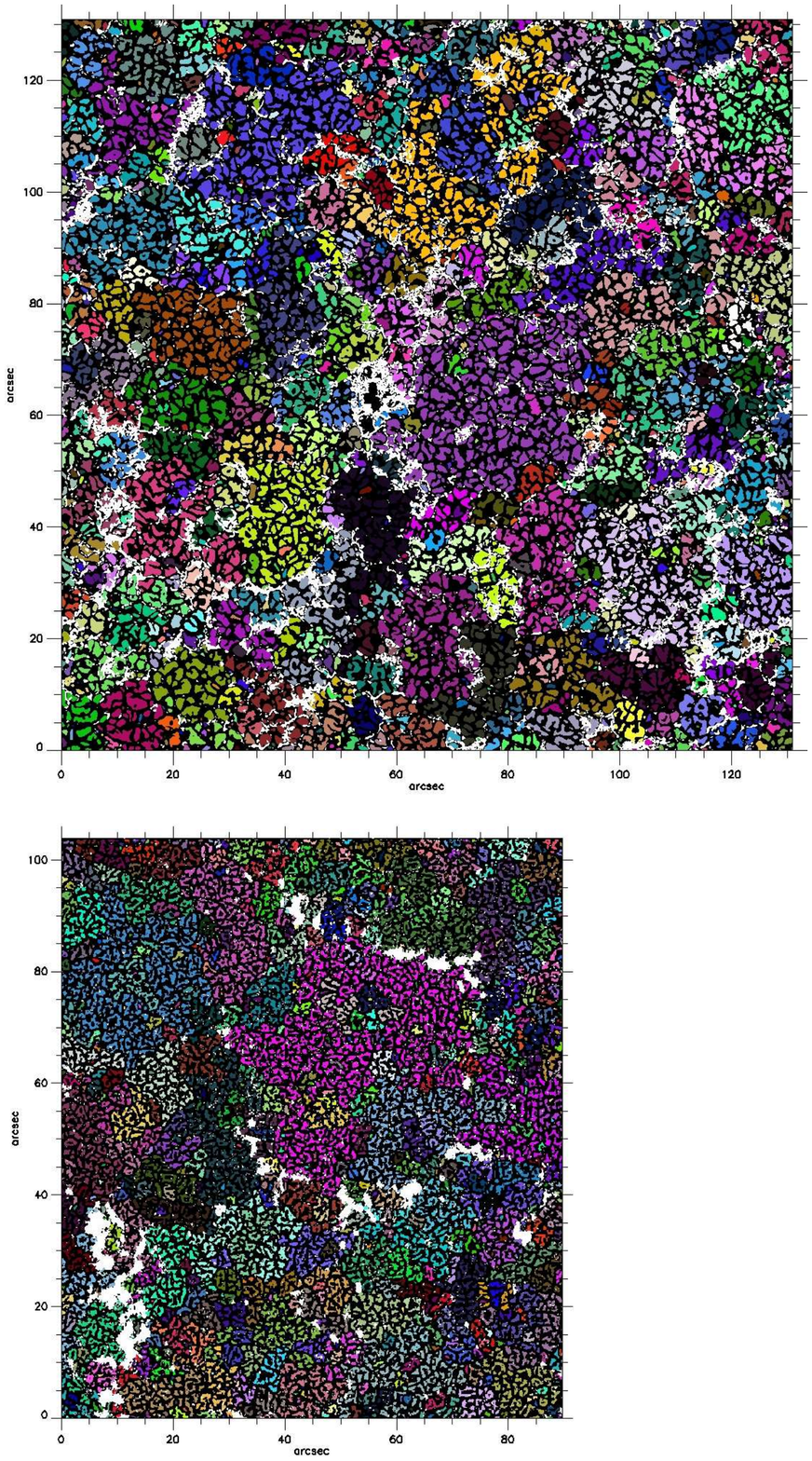}
 \caption{TGF at maximum extension (top: simulation; bottom: \textit{Hinode} 2007).
 Granules belonging to the same family have the same color.
 Magnetic fields (absolute value) are superimposed in white. } \label{fam}
\end{figure}

We discovered that TFG also form in the MHD simulation.
Figure~\ref{fam} displays TFG at the development time of the largest
family (30$''$ or supergranular size), with the absolute value of
BLOS superimposed. Magnetic fields are slowly advected to the
borders of families, in observations and simulation, while new
magnetic elements (located in intergranules) appear in the IN.

\begin{figure}
\centering
\includegraphics[width=8 cm]{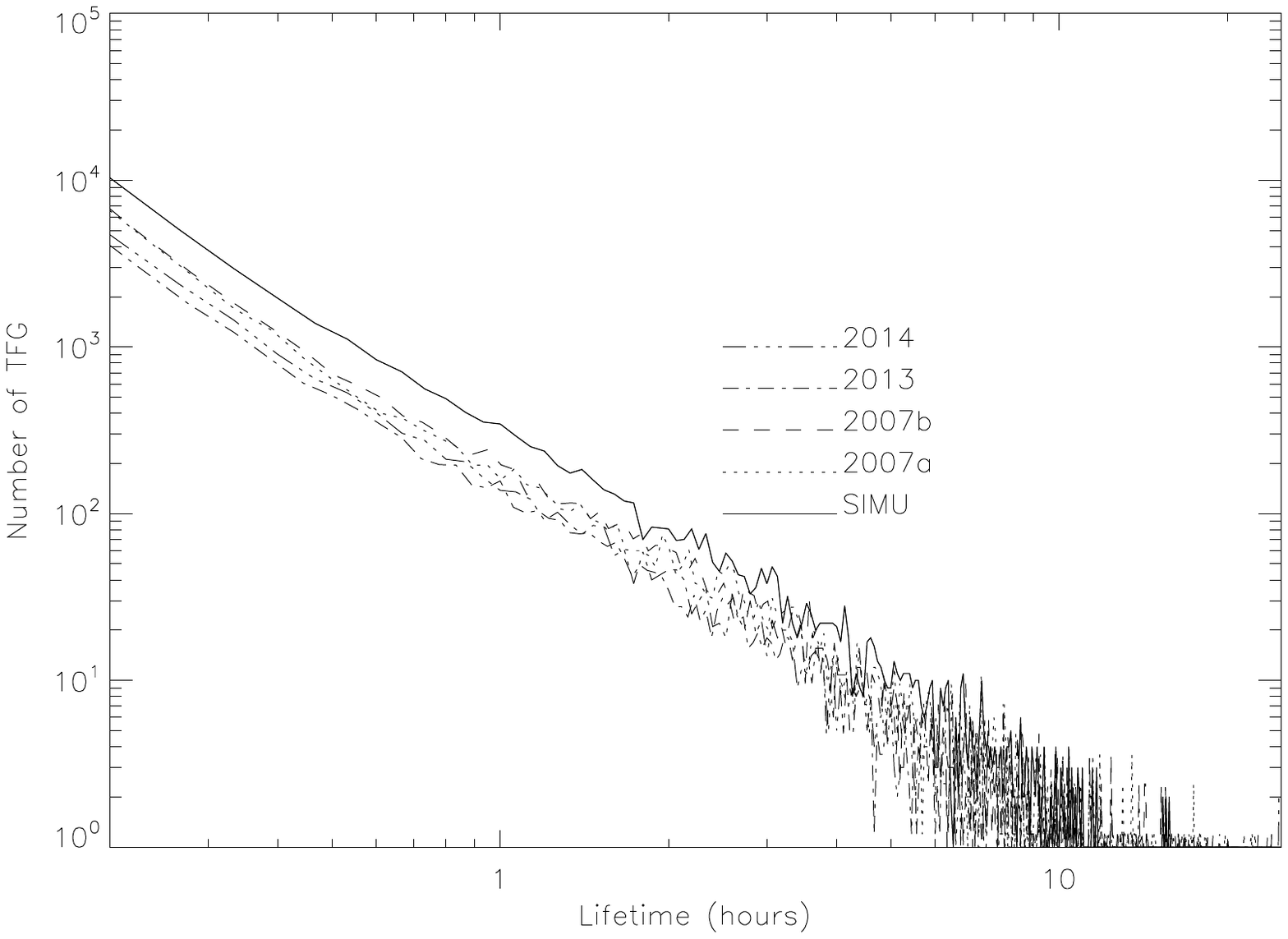}
 \caption{Distribution of TFG lifetimes: \textit{Hinode} observations
 of 2007, 2013 and 2014: dashed/dotted lines;
 simulation: solid line. } \label{duree}
\end{figure}

We now compare main characteristics of families using observations
(2007, 2013 and 2014) and numerical results. Figure~\ref{duree}
shows that the distribution function of family lifetimes is
typically a power law of the form $t^{-1.88}$, where $t$ is the
time. Most families develop to the mesoscale and have short lifetime
(a few hours), while some rare TFG may last 24 hours or more and
grow almost to the supergranular scale. The simulation appears in
remarkable agreement with \textit{Hinode}.

\begin{figure}
\centering
\includegraphics[width=8 cm]{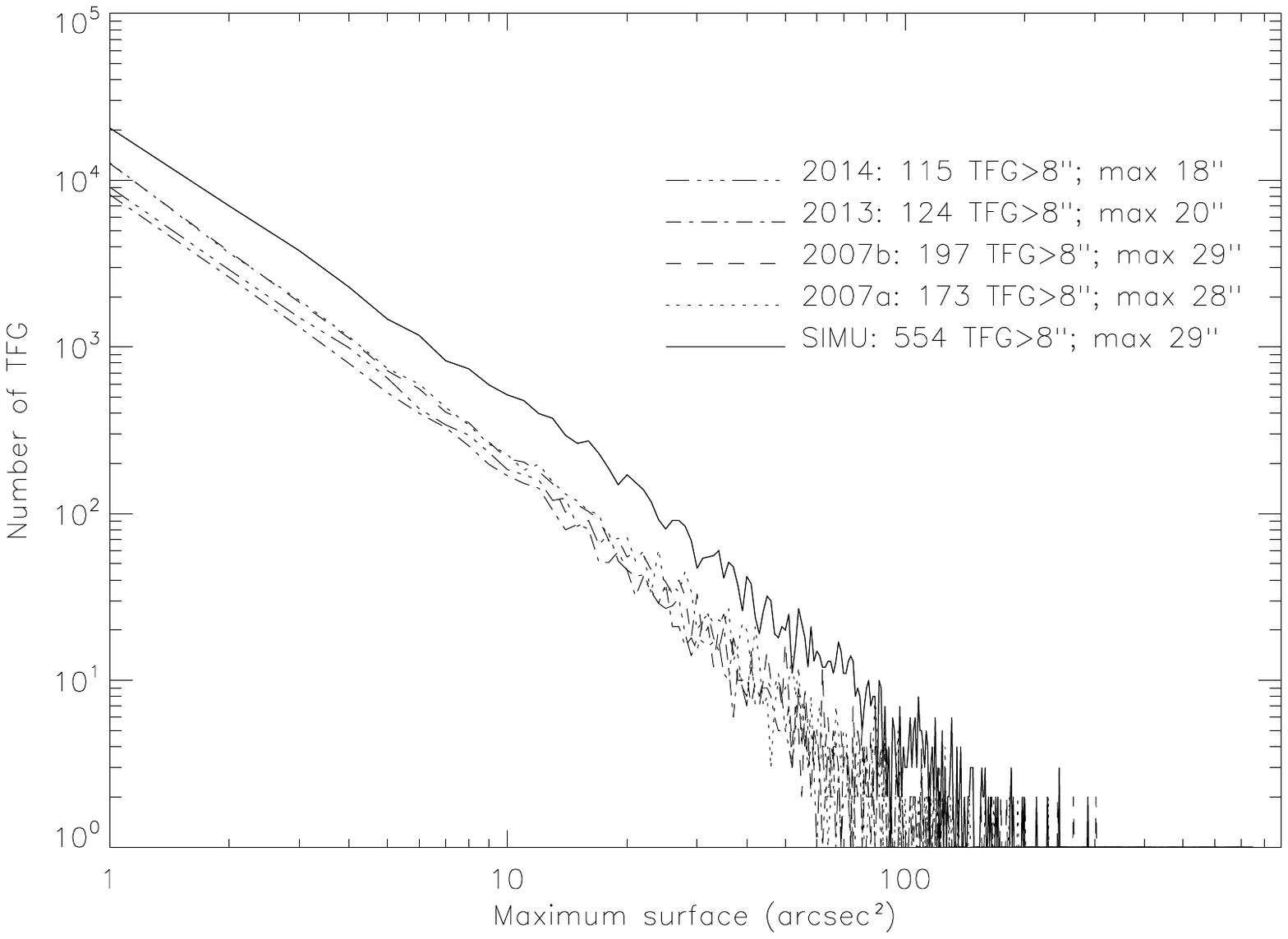}
 \caption{Distribution of TFG sizes: \textit{Hinode} observations
 of 2007, 2013 and 2014: dashed/dotted lines;
 simulation: solid line. } \label{taille}
\end{figure}

We also examined the distribution functions of the surface of
families, and found a typical power law of the form $t^{-1.74}$
(Figure~\ref{taille}). The maximum size corresponds to the
supergranular scale: 28-29$''$ for the simulation and
\textit{Hinode} 2007, 18-20$''$ for 2013 and 2014. However, large
families are not numerous: only ten families have areas in the range
500-900 $arcsec^{2}$ in the simulation; for 2007 observations, the
ten largest families are in the range 300-900 $arcsec^{2}$. Most
supergranules appear rather formed of several families at the
mesoscale (8$''$ typical size): in a unit area of 1$'$ $\times$
1$'$, we found 113 families in the simulation above 8$''$ and 80 in
observations (whatever the date).

\begin{figure}
\centering
\includegraphics[width=8 cm]{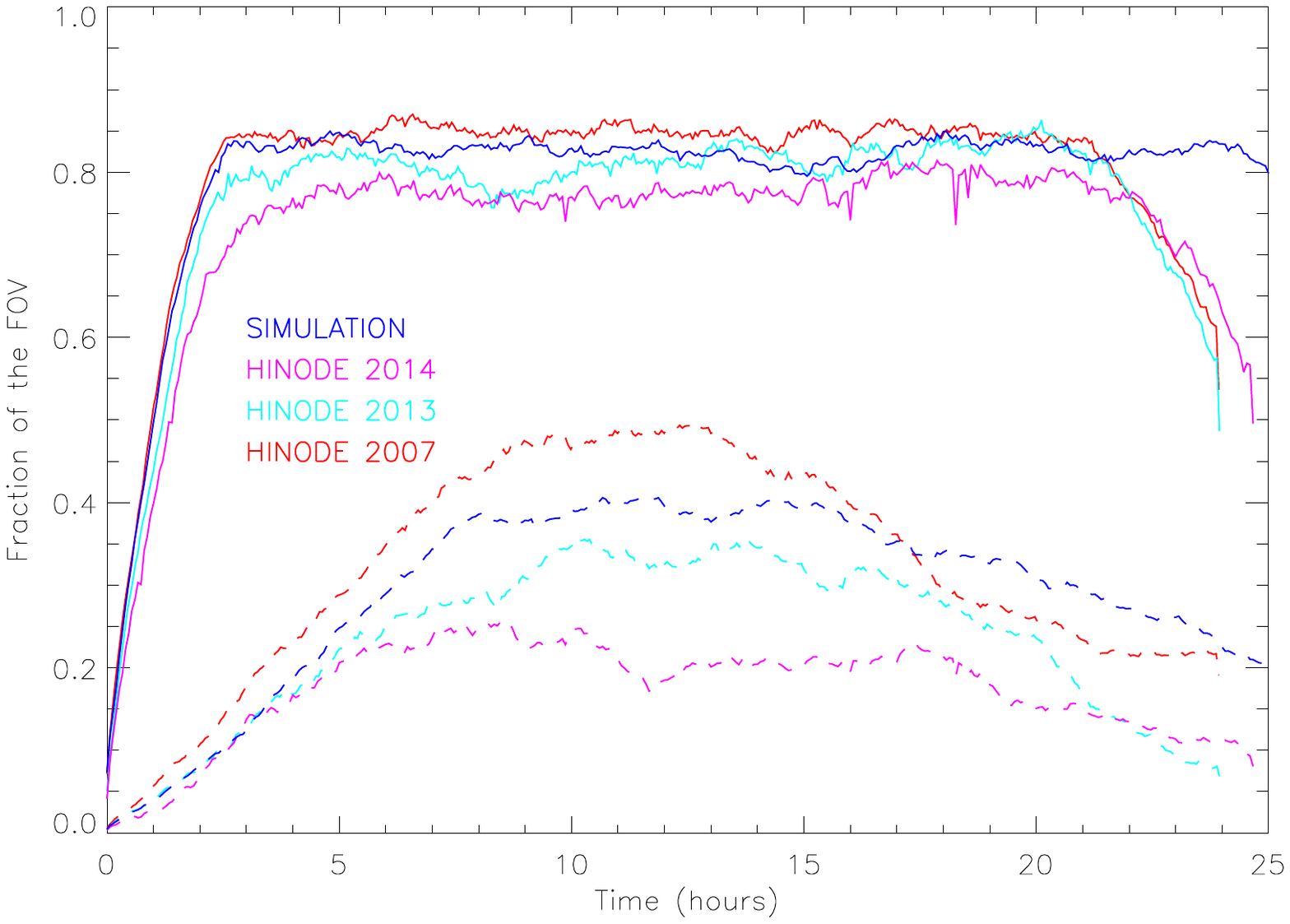}
 \caption{Fraction of the FOV covered by families lasting more than 3 hours (solid lines)
 or more than 12 hours (dashed lines): simulation (blue),
 2007 (red), 2013 (cyan) and 2014 observations (magenta).} \label{area}
\end{figure}

Figure~\ref{area} shows the contribution of families to the FOV for
two different lifetime thresholds as a function of time. Families
lasting at least 3 hours represent 75 to 85 \% of the FOV, while
long life families (lasting more than 12 hours) cover 20 to 45\% of
the solar surface (the dispersion is higher because the number of
long duration families is small).

Hence, we conclude that the spatio temporal properties of TFG as
seen by \textit{Hinode} or issued from the simulation fit well.

\begin{figure}
\centering
\includegraphics[width=8 cm]{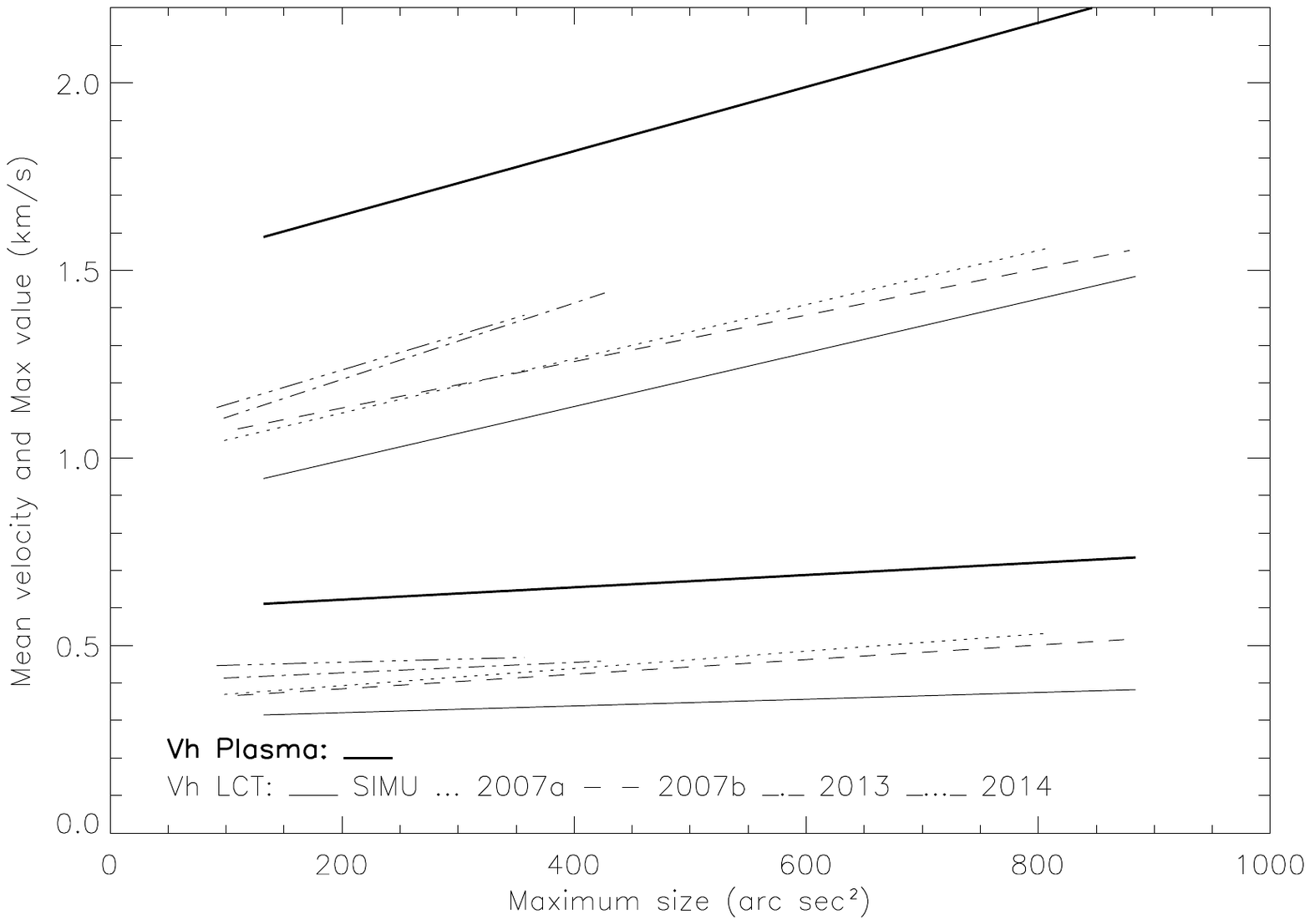}
 \caption{Mean (bottom) and maximum (top) horizontal velocities.
 Simulation: velocity plasma filtered by the SOT PSF (thick line) and LCT (thin line); observations:
 2007 (2 sequences, dashed and dotted), 2013 (dash dot) and 2014 (dash dot dot). } \label{dyna}
\end{figure}

We noticed that the largest TFG are the most dynamic. Mean and
maximum horizontal velocities of families are reported in
Figure~\ref{dyna} as a function of their maximum area. For
observations, velocities are derived from the LCT; for the
simulation, we used both LCT and plasma velocities (filtered by LCT
windows). While average velocities do not vary much with family
area, velocity maxima increase with size. We conclude that strongest
flows are generated by largest families composed of many
simultaneously exploding granules. Results are similar for
\textit{Hinode} and the MHD code.

Movies 1 and 2 display interactions between horizontal flows and
BLOS for 2007 observations and simulation. Velocity fronts
contribute to transport magnetic fields towards the boundaries of
supergranules delineated by the NE. Strongest fronts occur at the
border of large families, as shown in movies 3 and 4, suggesting
that the dynamics of largest (but not numerous) TFG could play an
important role in the NE buildup.

\begin{figure}
\centering
\includegraphics[width=8 cm]{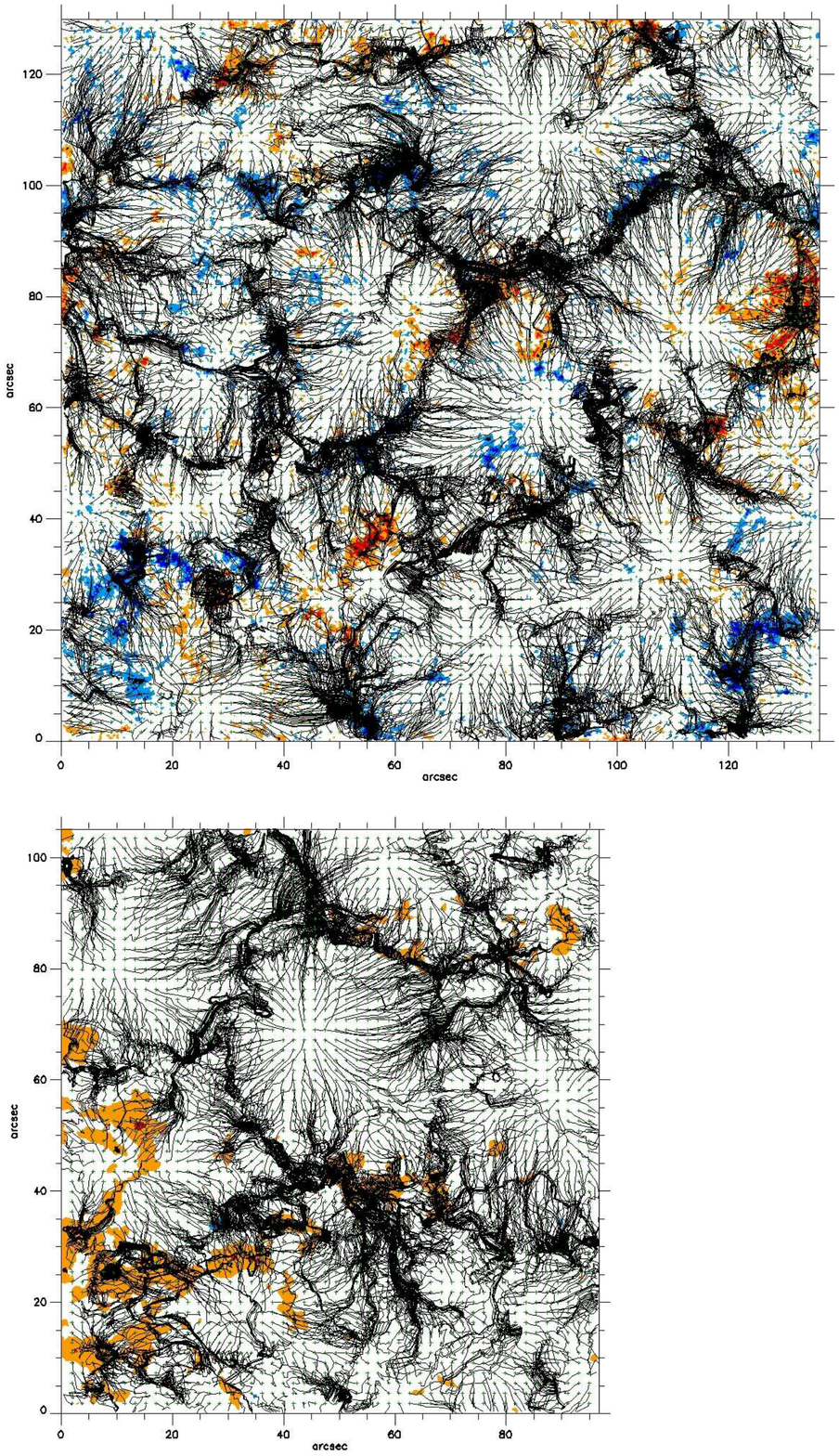}
 \caption{Trajectories of corks. Initial positions are indicated
 by green crosses. Magnetic polarities (green/blue and yellow/red)
 are superimposed. Top: simulation; bottom: \textit{Hinode} 2007}. \label{corks}
\end{figure}

The NE formation at supergranular scale is illustrated
(Figure~\ref{corks}) by the displacements of free corks, for
\textit{Hinode} 2007 as well as for the simulation. At time $t=0$,
corks were uniformly distributed on a regular grid. After initial
time, corks move at the horizontal velocity of the LCT, and their
trajectories are drawn. Final positions coincide with the magnetic
network and delineate the boundaries of supergranules. We found,
here again, an impressive agreement between observations and
simulation. Movies 5 and 6 display the density of corks as a
function of time, showing that corks are pushed towards the magnetic
NE (which also correspond to bright He \textsc{ii} 30.4 nm hot
structures of \textit{Hinode}/EIS 2007).

Cork motions can be characterized by a coefficient $\gamma$ assuming
that the square of the distance $d^{2}$ from the initial position to
the final position has the form of the power law $t^{\gamma}$. For
pure diffusion or random motion, $\gamma$ is about 1, but for
advective motions, $\gamma$ is higher. The LCT suppresses
systematically Brownian motion, so that it does not appear in
observations. On the contrary, this component is obvious in the
simulation and is superimposed to the shift towards the NE. We
computed the mean $\gamma$ coefficient of observations and
simulation. We found the following results.

\begin{enumerate}
  \item Simulation: plasma velocity at 0.13$''$, time step 60 seconds:
  $\gamma$ = 1.02 (Brownian motion dominates)
  \item Simulation: plasma velocity at 3.5$''$, time step 30
  minutes: $\gamma$ = 1.59 (advection dominates)
  \item LCT (3.5$''$, time step 30
  minutes):
  \begin{itemize}
    \item Simulation: $\gamma$ = 1.82
    \item BFI 2007: $\gamma$ = 1.66
    \item BFI 2013: $\gamma$ = 1.67
    \item BFI 2014: $\gamma$ = 1.64
  \end{itemize}
\end{enumerate}

The $\gamma$ values provided by the LCT for various datasets are in
good agreement. After a few hours, $\gamma$ decreases because corks
have formed the NE, as shown by movies 5 and 6 where the surface
density of corks is plotted. When corks reach cell boundaries, they
slowly drift along them and tend to collapse together at particular
points which correspond to intersections of several supergranules.
This phenemenon is more visible in observations (movie 5) than in
the simulation (movie 6) where the magnetic NE is more diffuse. Our
results can be related to those of Giannattasio et al (2014a and
2014b) based on NFI magnetograms in Na \textsc{i} D1 line: they
found 1.44 for magnetic element displacements and 1.55 for magnetic
pairs. However, corks are not magnetic elements; they move freely at
the speed of the plasma flow.

\section {Discussion}

The surface of the Sun is covered by solar granules which are
grouped in TFG or families. We used continuum intensities to
evidence and label exploding granules forming families. The BFI
provided intensities at 450 nm, but synthetic emergent intensities
are at 500 nm. Danilovic \emph{et al.} (2008) compared granulation
contrast seen by the \emph{Hinode} spectro polarimeter (SP) with MHD
simulation and found that, at 630 nm, the simulated contrast
decreases from 0.14 to 0.07 after applying the SOT PSF (close to the
observed value). Wedemeyer-B\"{o}hm and Rouppe van der Voort (2009)
compared BFI images in the blue (450 nm), green (555 nm) and red
(668 nm) continua to synthetic images degraded by the PSF and found
0.11 in the blue. In the present study, the BFI contrast at 450 nm
is 0.13. Granulation images of the simulation have a contrast of
0.155 at 500 nm, reducing to 0.105 after filtering by the SOT PSF.
We found that the recognition of TFG is little affected by the PSF
because granule evolution between two consecutive times relies on
the detection of a common surface. The PSF removes the small
granules or details but does not affect the formation of TFG, which
are built essentially by the middle and large size granules
(although some branches may be cut). This is why we discovered
recently (work still in progress) that TFG are still detected at the
SDO HMI resolution.

Averaged horizontal velocities were computed by the LCT through
windows of 30 minutes and 3.5$''$; using the simulation, we found a
good agreement between the LCT applied to continuum images and
plasma motion except that the velocity module is underestimated by a
factor two. Using another simulation, Verma \emph{et al.} (2013)
showed that the LCT recovers main features of the granulation
dynamics, but proper motions may be underestimated by factor of
three. Louis \emph{et al.} (2015) have also compared this technique
to simulated flow fields and concluded that LCT is a viable and fast
tool to retrieve velocities for large data sets, with good
correlation between vectors and underestimation of the magnitude.
Alternative methods are discussed by Welsch \emph{et al.} (2007).
Recently, Asensio Ramos \emph{et al.} (2017) developed a new
algorithm (DeepVel) working on consecutive images to evidence small
scale instantaneous motions, but providing similar results to LCT
for averaged velocity fields.

TFG sizes are distributed continuously with a decreasing slope from
small (mesoscale) to large (supergranule); this is also the case of
lifetimes. TFG compete each other but largest ones, which are not
numerous (power law), generate the strongest horizontal flows
pushing the IN magnetic field to the border of supergranules to form
the quiet NE. Thus, TFG appear as an essential part of the
supergranulation. This schematic view of TFG dynamics is compatible
with the evolution of the IN and maintenance of the NE described by
Go\v{s}i\`{c} \emph{et al.} (2014). They evidenced, using NFI
observations in Na \textsc{i} D1, a flux transfer from the IN to the
NE, supplying as much flux as present in the NE in 24 hours. We
report here large scale horizontal flows generated by the TFG
irrespective to the date along the solar cycle (2007, 2013 and 2014)
and suggest that their associated flows could contribute to the
transport of magnetic elements from the IN to the NE.

\section {Conclusion}

We have studied the dynamics of the quiet Sun from disk center
observations of the \textit{Hinode} SOT in terms of mesoscale
horizontal flows, evolution of TFG (trees of fragmenting granules or
families) and line of sight magnetic fields, along the cycle (2007
at solar minimum, 2013 and 2014 near solar maximum) using 24-hour
sequences; results were compared to those issued from the
magneto-convection code at similar space and time resolutions after
filtering by the SOT PSF. Horizontal flows in observations were
computed using the LCT applied to intensities; for the simulation,
we used both LCT and plasma velocity and found good agreement with
\textit{Hinode}. In all cases, TFG appear after a few hours, most at
the mesoscale, but some (composed of several branches) reach the
supergranular scale. Families are associated to velocity fronts
which advect magnetic fields and contribute to form the network.
Largest TFG are not numerous but the most dynamic; their development
could be an efficient mechanism to build the network. The simulation
provides realistic results about the properties, dynamics of TFG and
network interaction. However, small discrepancies do exist, as
granule merging/splitting rates or magnetic field polarities which
appear more mixed in the simulation than in observations (although
this may come from instrumental effects and remnant activity).

We do not see any striking variation of the dynamics between solar
minimum and maximum, confirming a previous study by Roudier \emph{et
al.} (2017) based on SDO/HMI observations. However, we need to
analyze more data between 2007 and 2013 in the ascending phase, but
unfortunately the number of exploitable sequences at disk center in
the blue continuum is rather limited. G band sequences are more
frequent but also more difficult to analyze with the LCT (bright
points). We checked that IRIS observations (slit jaw continuum at
283 nm) are LCT compatible, so that new data could be used in the
future to cover longer periods along the cycle. For that purpose,
HOP312 with IRIS has been set up.

\appendix

The Electronic Supplemental Material movies are are available in MP4
format.

\begin{itemize}

\item Movie 1: horizontal velocities ($\sqrt{v_{x}^{2}+v_{y}^{2}}$)
of \textit{Hinode}/BFI from LCT of blue continuum, 29-31 August
2007, sequence duration 48 hours, FOV $65'' \times 75''$. Velocities
in grey levels (LCT windows of 30 minutes/$3.5''$). The line of
sight magnetic field (Stokes \emph{V} as a proxy of BLOS from
\textit{Hinode}/NFI blue wing of Fe\textsc{i} 630.2 nm, pixel size
$0.16''$, 5 minutes averaged) is shown in blue/orange for
North/South polarities.

\item Movie 2: horizontal plasma velocities
averaged through 30 minutes/$3.5''$ filters (LCT windows for
comparison with movie 1) of the numerical simulation, sequence
duration 26 hours, FOV $131'' \times 131''$. Velocities in grey
levels. The vertical component of the magnetic field (pixel size
$0.13''$, 5 minutes averaged) is superimposed in blue/orange for
north/south polarities.

\item Movie 3: Families of granules (TFG in various colors) derived
from \textit{Hinode}/BFI blue continuum at 450.4 nm, 29-31 August
2007, sequence duration 48 hours, FOV $90'' \times 105''$, pixel
size $0.11''$, together with horizontal velocities from LCT
technique (30 minutes/$3.5''$ windows, grey levels).

\item Movie 4: Families of granules (TFG in various colors) provided
by the numerical simulation, sequence duration 25 hours, FOV $131''
\times 131''$, pixel size $0.13''$, together with horizontal plasma
velocities (grey levels, 30 minutes and $3.5''$ filtered through LCT
windows for comparison with movie 3).

\item Movie 5: density of corks (initially uniformly
distributed over the FOV) represented by disks (size proportional to
corks number, 10, 30, 100, 300, 1000, 3000 and 10000 or more). The
corks are driven by horizontal LCT velocities (30 minutes/$3.5''$
windows) from \textit{Hinode}/BFI blue continuum at 450.4 nm, 29-31
August 2007, sequence duration 48 hours, FOV $90'' \times 105''$.
BLOS (\textit{Hinode}/NFI blue wing of Fe\textsc{i} 630.2 nm) is
superimposed in blue/red for north/south polarities. In the
background, He\textsc{ii} 30.4 nm intensities from
\textit{Hinode}/EIS are displayed in green.

\item Movie 6: density of corks of simulation represented by disks
(size proportional to the number of corks). The corks are driven by
horizontal plasma velocities, sequence duration 26 hours, FOV $131''
\times 131''$. The vertical magnetic field is superimposed in
blue/orange for north/south polarities. The time step is 10 minutes
and the averaging window $3.5''$ (to allow comparison with movie 5).

\end{itemize}

\begin{acknowledgments}

  We are indebted to the \textit{Hinode} team for the possibility to
use their data. \textit{Hinode} is a Japanese mission developed and
launched by ISAS/JAXA, collaborating with NAOJ as a domestic
partner, NASA and STFC (UK) as international partners. Scientific
operation of the \textit{Hinode} mission is conducted by the
\textit{Hinode} science team organized at ISAS/JAXA. This team
mainly consists of scientists from institutes in the partner
countries. Support for the post-launch operation is provided by JAXA
and NAOJ (Japan), STFC (UK), NASA, ESA, and NSC (Norway).

The authors wish to acknowledge the anonymous referee and the editor
for helpful comments and suggestions to improve the manuscript.

Computing resources for the simulations were provided by the NASA
High-End Computing Program through the NASA Advanced Supercomputing
Division at the Ames Research Center.

This work was also supported by the Centre National de la Recherche
Scientifique, France. We acknowledge access to the HPC resources of
CALMIP under the allocation 2011-P1115.
\end{acknowledgments}

\end{document}